\documentclass[iop,apj]{emulateapj}
\slugcomment{ApJ in press}
\shorttitle{CMB-Regulated Star Formation}
\shortauthors{Bailin et~al.}

\newcommand{\ZCMB}{{\ensuremath{Z_{\mathrm{CMB}}}}}
\newcommand{\Zcrit}{{\ensuremath{Z_{\mathrm{crit}}}}}
\newcommand{\Tmin}{{\ensuremath{T_{\mathrm{min}}}}}
\newcommand{\TCMB}{{\ensuremath{T_{\mathrm{CMB}}}}}
\newcommand{\isotope}[2]{{\ensuremath{\mathrm{{}^{#1}#2}}}}
\newcommand{\overH}[1]{{\ensuremath{[\mathrm{#1}/\mathrm{H}]}}}

\begin{document}

\title{Consequences of cosmic microwave background-regulated star formation}

\author{Jeremy Bailin\altaffilmark{1}, Greg Stinson\altaffilmark{2},
  Hugh Couchman, William E. Harris, James Wadsley and Sijing Shen}
\affil{Department of Physics \& Astronomy, McMaster University,
  1280 Main Street West, Hamilton, ON, L8S 4M1, Canada}
\altaffiltext{1}{Current address: Department of Astronomy, University
  of Michigan, 500 Church Street, Ann Arbor, MI, 48109; jbailin@umich.edu}
\altaffiltext{2}{Current address: Jeremiah Horrocks Institute,
  University of Central Lancashire, Preston, PR1 2HE, United Kingdom}

\begin{abstract}
It has been hypothesized that the cosmic microwave background (CMB)
provides a temperature floor for collapsing protostars that can
regulate the process of star formation and result in a top-heavy
initial mass function at high metallicity and high redshift.
We examine whether this hypothesis has any testable
observational consequences.
First we determine, using a set of hydrodynamic galaxy
formation simulations, that the CMB temperature floor
would have influenced the majority of stars formed at redshifts
between $z=3$ and $6$, and probably even to higher redshift.
Five signatures of CMB-regulated
star formation are:
(1) a higher supernova rate than currently predicted at high redshift;
(2) a systematic discrepancy
between direct and indirect measurements of the high redshift
star formation rate; (3) a lack of surviving globular clusters
that formed at high metallicity and high redshift;
(4) a more rapid rise in the metallicity of cosmic gas than
is predicted by current simulations; and (5) an enhancement
in the abundances of $\alpha$ elements such as O and Mg
at metallicities $-2 \la \overH{Fe} \la -0.5$.
Observations are not presently able to either confirm or
rule out the presence of these signatures.
However, if correct, the top-heavy IMF of high-redshift
high-metallicity globular clusters could provide an explanation
for the observed bimodality of their metallicity distribution.
\end{abstract}


\section{Introduction}
The initial mass function (IMF) describes the relative numbers
of stars formed with different masses.
Observations suggest that the IMF has a power-law form at the high mass end
($M \ga 1~M_{\sun}$, with a slope near the canonical Salpeter value of
$\alpha=-2.35$; \citealp{salpeter-imf}), and a turnover at the
low-mass end \citep{kroupa01,chabrier03}.

The detailed form of the IMF is important
because stars of different
mass have different mass-to-light ratios, different lifetimes,
and different effects on their surroundings. Therefore, the
luminosity, chemical enrichment, and energetic feedback due
to a stellar population, as well as the evolution of these
quantities with time, all depend on the IMF.

In the local universe, observations indicate that the IMF is
universal, showing no evidence for variation between different
star formation events \citep{kroupa01}. Although we have no
full theory that explains the origin of the IMF
(see \citealp{mo07} for a good review),
it has recently been discovered that the functional form of the IMF is
the same as for the mass function of prestellar cores, but shifted to lower mass
\citep{all07}. This suggests that each star forms with
a constant fraction of its core mass,
and the functional form of the IMF is set by
the process of gas fragmenting into cores.
As such, it should be related to how the Jeans mass evolves
within a collapsing protostellar environment.
This leads to the intriguing possibility that star formation in
environments where the
thermal and density evolution is dramatically different than
in the local universe could result in different IMFs.

One such environment is the virtually metal-free gas that
formed the very first ``Population~III'' stars. Without any
metals, cooling below $10^4$~K is very inefficient, and
the fragmentation mass of the gas remains high. Detailed hydrodynamic
simulations of the formation of Population~III stars suggest
that stars formed from such zero-metallicity gas follow a very
top-heavy IMF, with a median
stellar mass perhaps as high as $100~M_{\sun}$ \citep{abn02}.
A more normal IMF is obtained when the gas reaches a critical
value \Zcrit, somewhere between $10^{-4}$ and $10^{-3}~Z_{\sun}$
\citep{bromm-etal01,ss07}.

Another important but less well-studied regime in which a top-heavy IMF has
been proposed is for high-redshift \textit{high}-metallicity gas.
Unlike primordial gas, which cannot cool efficiently enough to fragment,
high metallicity gas cools very efficiently. However, at high redshift the
cooling (and therefore fragmentation) comes to an abrupt halt when it reaches
the temperature of the cosmic microwave background
radiation (CMB).
Because it is thermodynamically impossible for the protostar to
cool below the CMB via radiative mechanisms, and cooling via adiabatic
expansion is not relevant to a collapsing protostar, the CMB
sets a temperature floor.
This abrupt halt to fragmentation results in a top-heavy IMF.
This argument was put forward in analytic form by \citet{larson05}, and was
further followed up numerically by \citet{omukai-etal05}, who used a single-zone
collapsing protostellar model with an advanced chemical network, and by
\citet{stson},
who performed \textit{ab initio} 3D hydrodynamic simulations
of high redshift star formation. These latter authors, in particular,
argue that high-redshift high-metallicity star formation must produce
very massive stars due to the influence of the CMB.

The idea that the CMB regulates star formation is intriguing
and could be very important. However, little
work has yet been done on estimating whether it would have had any
influence on real star formation, or on determining the observational
consequences of such regulation.
The goals of the present paper are to estimate the regimes
in which this effect could be important, and to examine possible consequences
that could be confirmed or ruled out by observations of the local
universe and at high redshift.
In \S~\ref{section:estimating-zcmb}, we provide a relationship
between the redshift of star formation and the critical
metallicity for CMB regulation.
In \S~\ref{section:zevolution}, we use high resolution cosmological
hydrodynamic simulations to estimate the amount of star formation
that occurred at different metallicities as a function of redshift.
We bring these together in \S~\ref{section:cmb-consequences} and
find that, if the CMB does couple to star-forming gas,
the majority of high-redshift star formation must have
happened in the CMB-regulated regime, and examine
the effects of this regulation on supernova and gamma ray
burst rates, direct vs. indirect measurements
of the high redshift star formation rate, the metallicity
distribution of old globular clusters, and the cosmic evolution
of both global metallicity and $\alpha$ abundances.
Finally, we present our conclusions in \S~\ref{section:conclusions}.

\section{Estimating \ZCMB}%
\label{section:estimating-zcmb}

According to the CMB-regulation hypothesis, star-forming gas that
is sufficiently metal-rich cools rapidly to the CMB temperature,
which halts fragmentation and results in a top-heavy IMF
\citep{larson05,stson}. The metallicity is a key
parameter because the cooling rate at $T<10^4$~K
is completely dominated by metal line emission for all but the
lowest metallicities \citep[e.g.][]{ssa08}. There should therefore be
at each redshift $z$
a critical metallicity, $\ZCMB(z)$, above which CMB regulation
is important.

\begin{figure}
\plotone{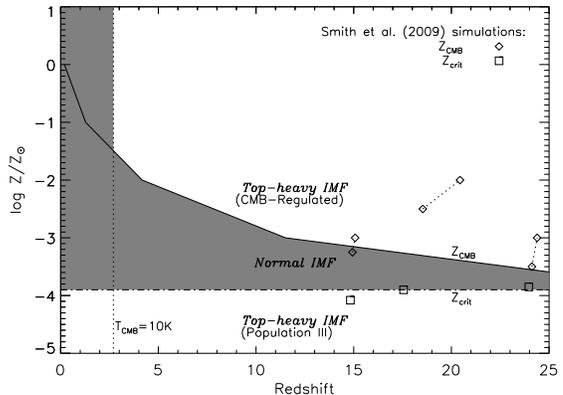}
\caption{\label{figure:zcmbest}%
Estimated evolution of \ZCMB\ as a function of redshift (solid line),
above which fragmentation is inhibited due to the CMB.
The dot-dashed line shows the critical metallicity \Zcrit, below
which fragmentation is inhibited due to inefficient cooling;
we use the value $\log Z/Z_{\sun}=-3.9$, from the Set~2
simulations of \citet{stson}.
The region in between these lines, where a normal IMF is expected,
is shaded. The dotted vertical line at $z=2.7$
denotes when the CMB temperature
drops below $10$~K; the CMB is not expected to be important below this
point. The results of \citet{stson} are also shown; the
squares denote the values of \Zcrit\ they calculated for each of their
sets of initial conditions,
while the pairs of connected diamonds denote the range
of possible values of \ZCMB.
Note that because the same initial conditions collapsed at different
redshifts for different metallicities, the pairs of diamonds are
not vertical.}
\end{figure}

The best estimates of \ZCMB\ to date come from
\citet{stson}, who performed \textit{ab initio} simulations with three
different sets of initial conditions and a variety of metallicities.
The amount of fragmentation they found in the simulations depended
on metallicity:
at low metallicities, no fragmentation was seen
because the gas remained hot;
at intermediate metallicities, the gas fragmented;
and at high metallicities, there was again no fragmentation
because the gas cooled so quickly that it reached the CMB temperature.
Because simulations can only be performed for a discrete set
of metallicities, we cannot determine the exact value of \ZCMB\ in
each case, but we can bound it on the lower end by the
most metal-rich simulation where fragmentation occured,
and on the upper end up the least metal-rich simulation
where fragmentation did not occur.
These are plotted in Figure~\ref{figure:zcmbest} as
the connected diamonds, at the redshifts at which each simulation 
collapsed.
Note that although the Set~2 simulations collapse at
a redshift intermediate between the Set~1 and Set~3 simulations,
their implied \ZCMB\ is higher, demonstrating that the boundary
between the normal and CMB-regulated regimes is not purely a function
of metallicity, but also depends on the details of the initial conditions;
\citet{jappsen-etal09} make a similar point regarding the lower metallicity
threshold, \Zcrit.

It is difficult to determine the consequences of CMB regulation
purely from these simulations because
very little star formation occurs at such high redshifts.
We must therefore devise
a method of estimating the evolution of \ZCMB\ as a function of redshift.
To do this, we note that
in the numerical models of \citet{omukai-etal05} and \citet{stson},
the CMB effectively acts as a temperature floor to the cooling collapsing
protostellar core. Without the CMB, the normal evolution of the core is
to cool and collapse until heat is released by
rapid $\mathrm{H_2}$ formation via three-body collisions,
which drives a rise in temperature.
Once all of the hydrogen is in molecular form, cooling
once again dominates and the temperature drops.
Eventually the core becomes optically thick, at which
point the dust emission can no longer cool the core,
and the temperature rises again due to compressional heating.
There is a minimum temperature $\Tmin(Z)$ that the gas reaches
during this evolution
(either immediately before the onset of the three-body reaction
or immediately before the core becomes optically thick),
which depends on its cooling history and therefore
its metallicity. Our ansatz is as follows: at
redshift $z$, the CMB regulates star
formation in gas of metallicity $Z$ if the CMB temperature
\begin{equation}
 \label{eqn:TCMB>Tmin}
 \TCMB(z) > \Tmin(Z),
\end{equation} where the CMB temperature at redshift $z$ is
\begin{equation}
  \TCMB(z) = T_0 (1 + z),
\end{equation}
for present-day CMB temperature $T_0 = 2.726$~K. If the effect of the
CMB is to act as a temperature floor, this condition defines
the regime where the CMB can have an effect.
We determine $\Tmin(Z)$ from the tracks of \citet{omukai-etal05}, who
modelled the evolution of temperature and density for a series of
metallicities $Z/Z_{\sun}=0$, $10^{-6}$, $10^{-5}$, $10^{-4}$, $10^{-3}$, 
$10^{-2}$, $10^{-1}$,  $10^0$. For each of these metallicities, we
can then determine the redshift where equation~(\ref{eqn:TCMB>Tmin})
is satisfied.

\begin{figure}
\plotone{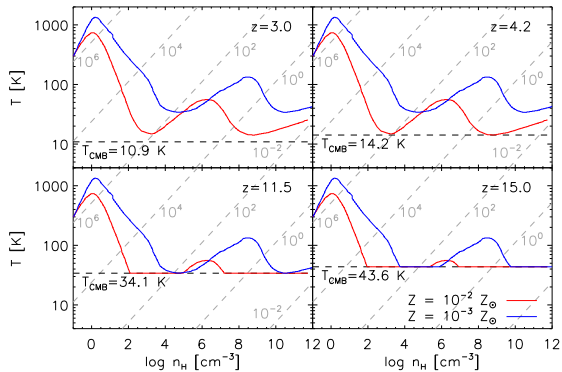}
\caption{\label{figure:omukai}%
Density-temperature evolution of a protostellar core with metallicity
$Z=10^{-2}~Z_{\sun}$ ($10^{-3}~Z_{\sun}$) in red (blue)
from \citet{omukai-etal05}, with a temperature floor set by
the CMB temperature at each of the four labelled redshifts.
The horizontal dashed lines denote the CMB temperature at each
redshift.
For $Z=10^{-2}~Z_{\sun}$, $\Tmin=14.2$~K, while
for $Z=10^{-3}~Z_{\sun}$, $\Tmin=34.1$~K.
The diagonal dashed gray lines indicate lines of constant
Jeans mass, and are labelled in $M_{\sun}$.}
\end{figure}

This is demonstrated in Figure~\ref{figure:omukai}, where we have
plotted the density-temperature evolution of protostellar cores
of two different metallicities ($\log Z/Z_{\sun} = -2$, $-3$)
from \citet{omukai-etal05}. In each of the four panels, we have imposed
a temperature floor given by \TCMB\ at a different redshift.
At $z=3$, $\TCMB=10.9$~K, which is colder than either core
ever reaches, and the CMB has no effect. At $z=4.2$,
$\TCMB=14.2$~K, which is the exact minimum temperature that the
$\log Z/Z_{\sun}=-2$ protostar reaches; if the CMB temperature were
any higher, it would affect the core evolution. This can be seen
at $z=11.5$, when $\TCMB=34.1$~K; the $\log Z/Z_{\sun}=-2$ core
does not cool nearly as far, and the CMB temperature just reaches
the minimum temperature of the $\log Z/Z_{\sun}=-3$ protostar.
At $z=15$, the temperature floor imposed by the CMB clearly affects
the evolution of both cores. We therefore adopt
$\ZCMB(z=4.2) = 10^{-2}~Z_{\sun}$ and
$\ZCMB(z=11.5) = 10^{-3}~Z_{\sun}$.
We perform the same calculation for each metallicity track
in \citet{omukai-etal05} and plot the relationship as the solid line
in Figure~\ref{figure:zcmbest}.

Fragmentation is only expected to be efficient when the Jeans mass
rapidly decreases, i.e. when the tracks move downward.
Therefore the
characteristic mass scale is the Jeans mass when the core stops
cooling \citep{larson05}.
Lines of constant Jeans mass are shown as the diagonal dashed gray
lines in Figure~\ref{figure:omukai}.
For the $Z=10^{-2}~Z_{\sun}$ track,
this characteristic scale is $0.2~M_{\sun}$, but at redshift $z=11.5$
it rises to $50~M_{\sun}$. The top-heavy IMF is a direct consequence
of this dramatic change in the characteristic mass scale at the
end of fragmentation.

Because the temperature of the protostellar core only just reaches
the CMB temperature, it is unlikely that the CMB has much effect
at the exact redshift we calculate. The $\TCMB(z)$, and therefore $z$,
that we calculate for each metallicity must be a slight underestimate, or
correspondingly the \ZCMB\ that we calculate at each redshift must
be a slight underestimate.

Our estimated \ZCMB\ matches the simulation results of \citet{stson}
very well for their Set~1 initial conditions, mildly underpredicts it
for their Set~3 intial conditions, and underpredicts it by about an
order of magnitude for their Set~2 initial conditions.
We should therefore expect that the transition between a normal
and top-heavy IMF occurs at metallicities
somewhere between our predicted \ZCMB\ and a metallicity
ten times larger, with the details depending on the
properties and formation history of the individual halos in
which the star formation occurs.

Although our estimate for \ZCMB\ is well-defined down to 
$z\approx 0.2$, it is
unlikely that CMB regulation is truly important at these redshifts.
As noted by \citet{stson}, gas clouds in the local ISM are not
observed to cool below $10$~K; therefore, when the CMB drops below
this temperature, it can no longer affect the thermal evolution of
protostellar gas. This occurs at $z=2.7$, denoted by the vertical
dotted line in Figure~\ref{figure:zcmbest}.

As discussed earlier, a top-heavy IMF is also expected for Population~III
stars, with metallicities below \Zcrit, denoted by the horizontal dot-dashed
line in Figure~\ref{figure:zcmbest}.
A normal IMF is therefore expected
in the shaded region of Figure~\ref{figure:zcmbest}; at lower
metallicities, cooling is too inefficient for fragmentation to occur,
while at higher metallicities, the gas cools immediately to the CMB
temperature where it becomes isothermal and does not fragment.
Although the estimates of the boundaries of the region are rough,
they provide a guideline for the star formation events that would
have been influenced by the CMB.

It is interesting that our derived \ZCMB\ rises substantially from
$< 10^{-3}~Z_{\sun}$ at $z>10$ to $\sim 10^{-1.5}$ at $z\sim 3$.
Such metallicities are typical for old stellar populations,
and it is therefore plausible
that CMB regulation could have been important for stars
formed at these redshifts.

\section{Cosmic Metallicity Evolution}
\label{section:zevolution}
In order to determine the effects of CMB-regulated star formation, we must
estimate, at each redshift $z$, the fraction of stars that formed with
metallicities $Z > \ZCMB(z)$.
Observational measurements are available at $z \la 3$ from
stellar population modelling of the integrated spectra of
galaxies from the Sloan Digital Sky Survey
(SDSS; \citealp{panter-etal08}),
but, as discussed in \S~\ref{section:estimating-zcmb}, the effects of the CMB
are only likely to be significant at $z \ga 3$. We must therefore use
theoretical models to estimate the cosmic evolution of the metallicity
of star-forming gas.

\subsection{Simulations}
\label{sec:sims}

Our simulations were performed as part of the McMaster Unbiased Galaxy
Simulations project (MUGS), a campaign to construct high resolution
simulations of a large set of $L^*$ galaxies that randomly sample
the sites of galaxy formation,
including a full range of modern galaxy formation physics.
Full details of the MUGS simulations will be presented
in Stinson et~al. (in preparation); we provide an overview
of the most important properties of the simulations below.

The simulations were performed using a WMAP~3 $\Lambda$CDM cosmology
with $H_0 = 73~\mathrm{km~s^{-1}~Mpc^{-1}}$, $\Omega_m=0.24$,
$\Omega_\Lambda=0.76$, $\Omega_{\mathrm{bary}}=0.04$, and $\sigma_8=0.76$
\citep{wmap3}.
Halos were chosen from a uniform-resolution $256^3$ dark matter-only
simulation in a box of side $50~h^{-1}$~Mpc
using the friends-of-friends algorithm \citep{defw85}.
A random selection of isolated halos with masses $4 \times 10^{11}~M_{\sun} \le
M \le 2 \times 10^{12}~M_{\sun}$ were chosen for resimulation at higher
resolution with full baryonic physics.
The highest resolution dark matter region, encompassing all matter
out to $5~r_{\mathrm{vir}}$, has particle mass
$m_{\mathrm{DM}}=1.1065 \times 10^6~M_{\sun}$, while gas
particles are placed inside $3~r_{\mathrm{vir}}$ and have initial masses
$m_{\mathrm{gas}}=2.2131 \times 10^5~M_{\sun}$.
There are typically $\sim 1000000$ gas and high resolution dark matter
particles each in the refined regions of the resimulations,
with the exact number depending on the halo mass and the geometry of the
Lagrangian region that collapses into the $z=0$ halo.

The simulations were evolved using the parallel SPH code \textsc{gasoline}
\citep{gasoline}. \textsc{gasoline} solves the equations of hydrodynamics
using SPH
and self-gravity using the Barnes-Hut tree algorithm \citep{bh86},
and includes radiative cooling, an ultraviolet (UV) background, star formation,
and energetic and chemical feedback.

The cooling is calculated from the contributions of both primordial gas
and metals as
$\Lambda_{\mathrm{tot}}(z, \rho, T, Z) = \Lambda_{\mathrm{HI, HeI, HeII}}(z,\rho, T) +
\frac{Z}{Z_{\sun}}\Lambda_{\mathrm{metal},Z_{\sun}}(z, \rho, T)$.
The first term employs atomic cooling based
on a gas with primordial composition heated by a uniform UV
ionizing background, adopted from
Haardt \&\ Madau \citep[in preparation; see][]{hm01} with
rates coefficient closely matching those cited in
\citet{Abel97},
while the metal cooling grid is constructed using Cloudy
(version 07.02, last described by \citet{CLOUDY}), assuming ionization
equilibrium, as described in \citet{sws09}.
The UV background is used in order to calculate the metal cooling rates
self-consistently. The cooling lookup table is linearly interpolated in
three dimensions (i.e., $\rho$, $z$, $T$) and scaled linearly with metallicity.

The star formation and feedback recipes are based on the ``blastwave model''
described in detail in \citet{Stinson06}, but with the addition of
clustered supernovae to account for the clustered nature of star formation.
Star formation can occur in gas particles that are dense
($n_{\rm min}=0.1~\mathrm{cm^{-3}}$) and cool
($T_{\rm max} = 15,000$~K),
calibrated to match the \citet{kennicutt98} Schmidt Law for the
Isolated Model Milky Way in \citet{Stinson06}.

At the resolution of these simulations, each star particle represents a
large number of stars ($6.32\times 10^4~M_{\sun}$). Thus, each particle has its
stars partitioned into mass bins based on the initial mass function presented
in \citet{Kroupa93}. These masses are correlated to stellar lifetimes as
described in \citet{Raiteri96}.
We stochastically determine when a star particle releases feedback
energy so that a minimum of $30$ supernovae worth of energy is
released concurrently to reflect the clustered nature of star formation.
The explosion of these stars is treated using the analytic
model for blastwaves presented in \citet{MO77} as described in detail in
\citet{Stinson06}. While the blast radius is calculated using the full energy
output of the supernova, less than half of that energy is transferred to the
surrounding ISM, $E_{SN}=4\times10^{50}$~ergs. The rest of the supernova energy
is radiated away. Iron and oxygen are produced in SNII according to the analytic
fits used in \citet{Raiteri96}.
The iron and oxygen are distributed to the same gas within the blast radius
as is the supernova energy ejected from SNII. Each SNIa produces $0.63~M_{\sun}$
iron and $0.13~M_{\sun}$ oxygen \citep{Thielemann86} and it is ejected into the nearest gas particle for SNIa.

We have implemented diffusion of all scalar SPH quantities,
particularly metal content and thermal energy, as described
in \citet{sws09}, which is required to correctly model even simple
processes such as convection and Rayleigh-Taylor instabilities
\citep{wadsley-etal08} and to account for mixing
in turbulent outflows.

MUGS galaxies are labelled by their group number in the list returned
by the friends-of-friends algorithm.
The simulations analyzed in this work are
MUGS~g$1536$, g$5664$, g$7124$, g$15784$, g$21647$, g$22437$, g$22795$,
and g$24334$. Only stars within the virial radius of the main galaxy
are considered.

\subsection{Metallicity Evolution}

\begin{figure}
\plotone{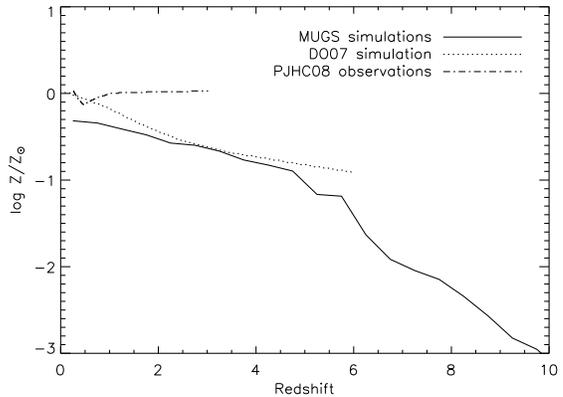}
\caption{\label{figure:zevol}%
Metallicity of stars as a function of their formation redshift. The solid line
is the mean from the MUGS simulations,
the dotted line is the mean from
the simulation of \citet{do07}, and the dot-dashed line is the
observed mean metallicity of star-forming gas at each redshift,
as determined from spectral synthesis modelling of SDSS
galaxies \citep{panter-etal08}.}
\end{figure}

The mean metallicity of stars formed in the simulation
is shown in Figure~\ref{figure:zevol} as a function of
their formation redshift.
The solid line shows the
results for all stars within MUGS simulated galaxies, and shows that
stellar metallicities rise from $\sim -3$, for those formed
at $z \ga 10$, to nearly solar for those formed at $z=0$.
Our simulations are of $L^*$ galaxies, while the majority of
star formation at high redshift occurred in larger galaxies, which
formed their metals earlier than less massive galaxies.
Our determinations may therefore underestimate the metallicities
of a universal sample of stars formed at high redshift.

We have confirmed that our results are consistent with those
obtained with completely different codes:
the dotted line, which shows the
star-formation-rate-weighted mean metallicity of gas from the
GADGET2 simulations of \citet{do07} (the ``SFR-weighted'' line in
their figure~2) and should be directly comparable,
shows reasonable agreement with our results
over the entire range $0 \le z \le 6$ that they plotted.
Where our results deviate from those of \citet{do07}, it is
in the sense that the MUGS metallicities are lower.

The best observational measurements of stellar metallicity as a function
of formation redshift come from \citet{panter-etal08}, who performed spectral
synthesis modelling of SDSS galaxies at
redshifts ranging from $0.1$ to $3$. The mean metallicities of
stars inferred to have formed at each redshift is shown
as the dot-dashed line in Figure~\ref{figure:zevol}. Unlike
the simulation predictions, the observations show essentially no
drop in stellar metallicity out to $z=3$. This is mainly because
the total stellar mass is dominated by the most massive galaxies,
which formed most of their stars very early; however, the stellar
metallicities
within galaxies of stellar mass $3 \times 10^{10}~M_{\sun} \le
M_* \le 1 \times 10^{11}~M_{\sun}$ show almost identical behavior,
and this is very nearly the same mass range as the MUGS galaxies
($4 \times 10^{10}~M_{\sun} < M_* < 1.2 \times 10^{11}~M_{\sun}$),
and the mass range that includes the Milky Way
($M_* \approx 5 \times 10^{10}~M_{\sun}$; \citealp{kzs02,wpd08}).
We may therefore conclude that our simulations do not overpredict
the metallicity of star-forming gas at high redshifts, and may
even underpredict it, an issue we will return to in \S~\ref{section:abundances}.

\section{Consequences of CMB regulation}%
\label{section:cmb-consequences}

\subsection{When was the CMB important?}

\begin{figure}
\plotone{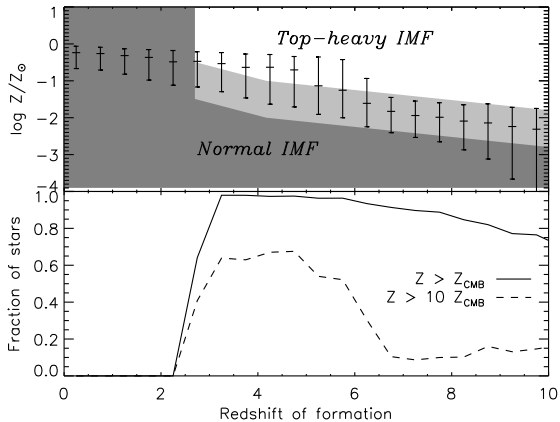}
\caption{\label{figure:mugszcmb}\label{figure:fracabove}%
\textit{(Top)}
The data points denote the median metallicity of stars formed at
each redshift within the MUGS simulations, while the error
bars denote the 10th and 90th percentiles.
Stars formed within the dark shaded region, which is identical
to the shaded region in Figure~\ref{figure:zcmbest}, are expected
to form with a normal IMF.  The light
shaded region extends to metallicities $1$ dex higher than \ZCMB,
where stars may form with either a normal or top-heavy IMF,
if CMB regulation occurs. All stars formed in the white region
are expected to exhibit a top-heavy IMF if the CMB regulation
hypothesis is correct.
\textit{(Bottom)}
Fraction of stars formed with metallicities greater than \ZCMB, i.e. within
the light shaded or white regions of the top panel. The dashed line
denotes the fraction of stars formed with metallicities greater than
$10~\ZCMB$, i.e. within the white region of the top panel.}
\end{figure}

We have plotted our estimate of the metallicity of star-forming
gas from \S~\ref{section:zevolution} over our estimate of the
critical metallicity for CMB regulation, \ZCMB, from
\S~\ref{section:estimating-zcmb},
in the top panel of Figure~\ref{figure:mugszcmb}.
Note that unlike in Figure~\ref{figure:zevol},
here we plot the \textit{median} metallicity, along with the 10th and
90th percentiles.

The first remarkable thing to note about Figure~\ref{figure:mugszcmb}
is that the median
star was formed with $Z>\ZCMB$ at almost all redshifts greater than $2.7$.
In other words, if the CMB regulation hypothesis is correct,
\textit{the majority of stars at high redshift were formed
in the CMB-regulated regime}. While this statement is subject to 
uncertainties in both our estimate of \ZCMB\ and
in the cosmic metallicity evolution measured from the simulations,
it would appear that the CMB influenced a significant amount of star formation
at high redshift.

In the bottom panel of
Figure~\ref{figure:fracabove}, we have plotted the fraction of stars that
form at $Z>\ZCMB$ in the simulation. As expected from the top panel,
this is very high: over $80\%$ at virtually
all redshifts where the CMB temperature
is higher than $10$~K. Given the uncertainties in both the estimate of
\ZCMB\ and in the metallicity evolution predictions of the simulation,
we have also determined the fraction of stars formed with metallicity
$Z > 10~\ZCMB$ (dashed line).
We consider it unlikely that
the combined difference between the simulation metallicities and the \ZCMB\ 
estimate is a full order of magnitude;
moreover, Figure~\ref{figure:zevol} strongly suggests that the simulated
metallicities are more likely underestimates than overestimates, which
would imply that an even larger fraction of stars formed with
$Z > \ZCMB$ or $Z > 10~\ZCMB$ than has been calculated.
However, even with this drastic change, half of the stars
formed at $2.7 < z < 6$ are
formed in the CMB-regulated regime, a fraction that drops to $10$--$15\%$
at higher redshift.

Our first major conclusion is therefore that if the CMB regulation
hypothesis is correct, the CMB must have influenced the majority
of star formation at $3 \la z \la 6$, and may have been important
to even higher redshift.

\subsection{Supernova and Gamma Ray Burst Rates}
\label{section:snrate}
Stars whose initial masses are at least
$M_{\mathrm{SN}} \sim 8~M_{\sun}$ end their
lives as core-collapse supernovae. The fraction of stars above
this limit, and therefore the number of supernovae produced per
unit mass of stars depends on the IMF: a top-heavy IMF produces
more supernovae. For a truncated power-law IMF, this ratio is
given by
\begin{equation}
\frac{N_{\mathrm{SN}}}{M_*} = \frac{(\alpha+2) (M_{\mathrm{max}}{}^{\alpha+1}
  - M_{\mathrm{SN}}{}^{\alpha+1})} {(\alpha+1) (M_{\mathrm{max}}{}^{\alpha+2}
  - M_{\mathrm{min}}{}^{\alpha+2})}
\end{equation}
\citep{bailinharris09-gcenrichment}.
The IMF may be top-heavy in the sense of having a flatter slope $\alpha$,
or be ``bottom-light'' in the sense of having a higher minimum stellar
mass $M_{\mathrm{min}}$. To give an example of the quantitative size
of the effect, flattening the slope
from the canonical Salpeter value of $\alpha=-2.35$ to $\alpha=-2.1$
increases the supernova rate by $50\%$, while increasing
$M_{\mathrm{min}}$ from $0.30~M_{\sun}$ to $5~M_{\sun}$
increases the supernova rate by a factor of $3.5$.
If the CMB-regulation hypothesis is correct,
and the majority of star formation at $3 \la z \la 6$ occurred
with a top-heavy IMF, then
predictions of the high-redshift supernova rate \citep[e.g.][]{df99}
are underestimates by a corresponding factor. Some indirect consequences
of this are discussed in sections~\ref{section:sfh-recon} and
\ref{section:abundances}.

The link between core-collapse supernovae and long duration gamma
ray bursts (GRBs) is now well established \citep{wb06}.
One might therefore expect that a top-heavy IMF,
which increases the supernova rate, might
also increase the GRB production rate.
However, there is both
observational and theoretical evidence that GRBs are preferentially
produced by metal-poor stars \citep[e.g.][]{yln06,wp07}. As CMB
regulation results in a top-heavy IMF only at high metallicities,
it is not obvious whether it would result in an enhancement of the GRB rate
without more detailed modelling.

\subsection{Reconstructing the cosmic star formation history}
\label{section:sfh-recon}
A key goal in the study of galaxy formation is 
to reconstruct the cosmic
star formation history \citep[e.g.][]{lilly-etal96,madau-plot}.
There are three
types of methods for performing this measurement:
the first is to directly
measure indicators of current star formation (such as H$\alpha$, UV,
far-infrared, or radio continuum emission)
in galaxies at a variety of redshifts;
the second is to deconstruct the stellar populations of low-redshift
galaxies to determine their ages; and the third is to examine the stellar
mass density in galaxies at a variety of redshifts, which is the
integral of the star formation rate minus the stellar death rate
over time.
The first of these methods is \textit{direct}: it measures
star formation as it happens, while the latter two methods are
\textit{indirect} and rely on the stars that are produced.

A key assumption in the indirect methods is the IMF: the mass of
stars remaining after a given length of time from a fixed burst
of star formation depends strongly on their mass distribution.
If the IMF is very top-heavy, then the number of extant stars
a given length of time later will be smaller than if the IMF is
normal. In the extreme case where all stars are massive, then
a burst of star formation will leave no visible stellar population
a short time after the burst, and will be completely missed by
any method that relies on the stars to measure the rate of star
formation.

Some of the direct star formation rate indicators also depend
on the IMF. For example, most radio continuum emission in star-forming
galaxies is synchrotron radiation from relativistic electrons that
are accelerated in supernova remnants \citep{condon92}. As discussed in
\S~\ref{section:snrate}, a top-heavy IMF increases the supernova rate
with respect to the star formation rate, and therefore must also
increase the radio continuum emission.
Therefore, a top-heavy IMF both increases the derived star formation
rate from direct indicators, and decreases the derived rate from
indirect indicators.

In fact, these measurement methods do not agree. At high redshift,
the direct measurements of the star formation rate are systematically
higher than the indirect methods.
This can be seen in \citet{heavens-etal04}, who perform
spectral synthesis modelling of SDSS galaxies at low redshift
to determine their stellar populations, and use them to
infer the star formation rate at redshifts up to $10$.
They find that their star formation rates are systematically
lower than the direct measurements in their higher redshift bins,
precisely where CMB regulation would be expected to be important.

A similar result comes from \citet{wth08}, who used measurements
of the stellar mass density as a function of redshift. The time derivative
of this is the net change in stellar mass, which is the rate of
star formation minus the rate that stars die. The latter term is
a strong function of the IMF, and is much higher for a top-heavy IMF,
requiring a much larger star formation rate to fit the same data.
These authors find that their implied star formation history is
consistent with direct measurements out to $z=0.7$, but
becomes increasingly discrepant at higher redshift, reaching a factor
of $4$ by $z=3$. They suggest that an increasingly top-heavy IMF
could explain this discrepancy, but do not offer a physical
mechanism for this evolution.
Although CMB regulation offers a natural explanation
for why the IMF would change with time,
one problem is that \citet{wth08} find
evidence for a discrepancy between the direct and indirect
star formation rates down to redshifts of $z=0.7$, while CMB
regulation is unlikely to operate much below $z=2.7$.
However, the discrepancy is relatively mild between $z=0.7$ and $z=2$,
and it is only beyond $z=2$ that the differences become significant.

Similar considerations drove \citet{dave08} to suggest an average IMF
that slowly evolves with time; although he mentions the influence
of the CMB as a potential driver for this evolution, the model is
expressed as an overall gradual change with redshift rather than
a change that specifically operates in the high-metallicity high-redshift
regime. \citet{fardal-etal07} perform a similar exercise while also considering
the energy contained in the extragalactic background light (which is
dominated by light from stars formed at $z<2.6$) as a constraint,
and find that the best way to reconcile the measurements is with a
top-heavy IMF, which they suggest occurs in
the starbursting galaxies that are increasingly common at redshifts
beyond where the extragalactic background is generated. A CMB-regulated
top-heavy IMF would perform a very similar function.

\subsection{Globular Cluster Bimodality}%
\label{section:gcs}

A conundrum in the study of globular cluster systems around
galaxies is the observed bimodality of their metallicity distribution
\citep[e.g.][]{harris-etal06}.
A possible explanation is that metal-poor clusters
formed during the initial collapse of the protogalaxy, while the
metal-rich population formed during subsequent major mergers. However,
the metal-poor peak covers a narrow range in metallicity,
which appears to be impossible to reproduce unless the formation
of metal-poor clusters was abruptly truncated \citep{beasley-etal02}.

However, if CMB-regulated star formation results in mostly
or exclusively massive stars, there is another possibility:
the first round of globular cluster formation did occur over
a wide range of metallicities from $\log Z/Z_{\sun} \sim -2.5$
upward, but the ones with $\log Z/Z_{\sun} \ga -1$ would have formed
with a top-heavy IMF.
The massive stars in the cluster would have died quickly, shedding the
majority of their mass in the form of stellar winds and supernova ejecta.
If the massive stars made up a significant fraction of the cluster mass,
as they would if the IMF were sufficiently top-heavy, this dramatic loss
of binding mass would result in the gravitational unbinding and dispersion
of the cluster.

In other words, the sharpness of the low-metallicity peak
and the relative absence of intermediate-metallicity clusters
near $\log Z/Z_{\sun} \sim -1$
may be entirely an artifact of a top-heavy IMF:
higher metallicity clusters formed along with their
lower-metallicity brethren, but did not survive to the present day.

\begin{figure}
\plotone{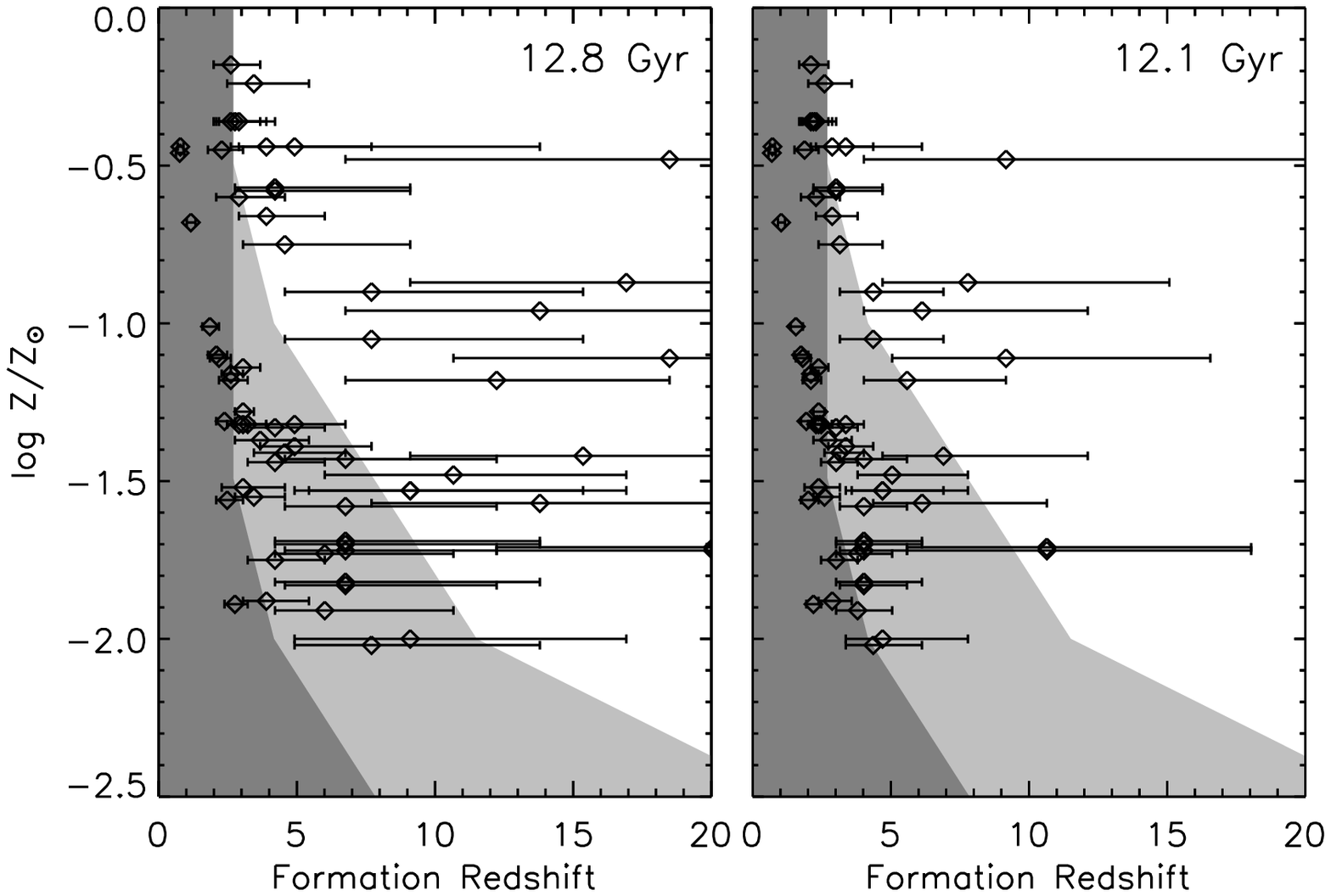}
\caption{\label{figure:gcs}%
Metallicity versus formation age of Milky Way globular clusters.
Data come from \citet{marinfranch-etal09}.
The relative ages have been converted into formation redshift
by assuming (1) the WMAP~5 recommended cosmological parameter values
\citep{wmap5}, and
(2) an absolute age scale of $12.8$~Gyr and $12.1$~Gyr for the
left and right panels respectively.
The dark shaded region is the 
region expected to exhibit a normal IMF, and is identical
to the shaded region in Figure~\ref{figure:zcmbest}. The light
shaded region also includes metallicities up to $1$ dex larger,
as in Figure~\ref{figure:mugszcmb}.}
\end{figure}

In the left panel of Figure~\ref{figure:gcs},
we have repeated the shaded and light shaded regions from
Figure~\ref{figure:mugszcmb} that denote the range in
redshift and metallicity where a normal IMF would be expected,
while the unshaded regions are those where CMB regulation,
if it occurs, would result in a top-heavy IMF.
We have overplotted the formation
redshifts and metallicities of Milky Way globular clusters
from the main sequence fitting performed by \citet{marinfranch-etal09}.
We have used the ages based on the \citet{dotter-etal07} isochrones,
and assumed that a normalized age of $1$ corresponds to an absolute
age of $12.8$~Gyr, consistent with the \citet{dotter-etal07} models.
These absolute
ages have been converted to formation redshift by assuming the
WMAP~5 recommended cosmological parameter values including
the results of WMAP, baryon accoustic oscillations, and type Ia
supernovae \citep[the WMAP+BAO+SN set of][]{wmap5}.
While the relative ages are precise to within $2\%$--$7\%$,
at high redshift a very small variation in age corresponds to a
large variation in redshift, and so the horizontal error bars
in Figure~\ref{figure:gcs} are very large at high redshift.
We have assumed the \citet{zw84} metallicity scale.

Examination of the left-hand panel of Figure~\ref{figure:gcs}
reveals that there are many GCs that have metallicities significantly
larger than the \ZCMB\ at which they formed,
i.e. above the dark shaded region, even given the large
uncertainties in the formation redshifts due to the finite precision
of the relative ages.
However, if we assume that normal star formation can occur at metallicities
up to $1$ dex higher than \ZCMB, i.e. in both the dark and light
shaded regions, then the vast majority of GCs lie within
the permitted region and only a few spill over into the forbidden region;
the most discrepant clusters are
Lyng\aa~7, NGC~6171, NGC~6717, and NGC~6362. Are the existence of these
clusters then conclusive evidence that CMB regulation is not
important?

It is interesting that three of these four clusters, NGC~6171,
NGC~6362, and NGC~6717,
lie in the less populated $\log Z/Z_{\sun} \sim -1$ metallicity regime between the
metal-poor and metal-rich peaks, and that all of these clusters
are quite low mass. Perhaps these objects are
incompletely-disrupted remnants of a larger population of clusters
that once inhabited the same parameter space,
or perhaps the transition from a normal IMF to a strongly top-heavy
IMF is gradual and these clusters formed with only a slightly top-heavy IMF.
In these cases, we might expect to see systematic differences between
the stellar mass functions of these clusters 
compared to those of other Galactic GCs;
detailed luminosity functions of these clusters would be well
worth measuring to assess the viability of this explanation.

However, we caution against overinterpretation of
the location of these clusters.
A serious concern is the ages:
while the relative ages of clusters are precise,
there is still large systematic uncertainty in the absolute
age scale. The effects of this can be seen clearly by comparing
the left-hand panel of Figure~\ref{figure:gcs} to the right-hand
panel, where we have reduced the absolute age scaling by $5\%$ to
$12.1$~Gyr. With this small change, the majority of GCs are
consistent with forming at metallicities below \ZCMB, and the
only cluster that is clearly in the forbidden region is
Lyng\aa~7. The age of this particular cluster is suspect,
however; other determinations of its age make it much younger
\citep{obb93,sarajedini04}, and its location in the bulge
introduces large uncertainties due to significant reddening.

Because of the large inherent formation redshift uncertainties,
we are not able to come to any firm conclusions regarding whether
CMB-regulated star formation has shaped the population of
extant globular clusters, although they appear to be consistent
with the CMB-regulation hypothesis if the few discrepant clusters
are indeed remnants of a disrupted population or simply
observational spill over.
Perhaps if more precise absolute ages
of GCs can be determined in the future,
or if the detailed luminosity functions of the interesting
clusters can be measured,
it may be possible
to address this issue more conclusively.

\subsection{Abundances and Abundance Patterns}%
\label{section:abundances}

A top-heavy IMF in metal-rich high-metallicity star formation may
leave characteristic patterns in the evolution of both the global
metallicity and the abundances of particular elements.
Not only does a top-heavy IMF increase the number of supernovae
produced, but it also increases the average mass of the supernova
progenitors\footnote{At least, for an IMF that is truly top-heavy
in the sense of having a flatter high-mass slope. 
This is not true of a ``bottom-light''
IMF, which is simply missing some fraction of low-mass stars but has
otherwise the same functional form.}.
More massive stars eject a larger fraction of their mass in the form
of newly-synthesized metals, and are particularly efficient
at forming $\alpha$ elements such as oxygen, neon and magnesium
\citep[e.g.][]{ww95}.

In fact, there is a slight inconsistency in using our simulations,
which assume a normal IMF, to predict the fraction of star formation
that occurs with a top-heavy IMF, as in \S~\ref{section:zevolution}.
If a large fraction of high-redshift star formation occurred in the
CMB-regulated regime, as we predict, then they must have ejected
a larger mass of metals into their environment, and therefore the global
metallicity of the universe must have risen more steeply when CMB
regulation was important, between at least $z=6$ and $z=3$, than the simulations
predict. This would result in an even larger fraction of high-redshift
star formation occuring in the CMB-regulated regime; however, we already
predict that the majority of high-redshift star formation occured above
\ZCMB, so this inconsistency does not qualitatively affect our conclusions.

However, this more rapid rise in the global metallicity is precisely
what is required to reconcile the discrepancy between the metallicities
predicted by the simulations, $\log Z/Z_{\sun} \sim -0.5$ at $z=3$, and
those observed, which reach solar metallicity by the same redshift
(see Figure~\ref{figure:zevol}).

A top-heavy IMF may also leave an imprint on the abundance patterns.
Core collapse supernovae from high mass 
progenitors eject a larger fraction of their mass in
the form of $\alpha$-elements, especially \isotope{16}{O},
but also \isotope{20}{Ne}, and \isotope{24}{Mg}
\citep{ww95,nomoto-etal97}. Therefore, a top-heavy IMF may
produce more $\alpha$ elements than a population with a
normal IMF.

Current Galactic chemical evolution models accurately reproduce
the observed abundance patterns of Galactic stars over a large range of
metallicities, leaving little room for significant changes. However,
one element whose abundances are not well reproduced is Mg,
which is systematically underpredicted by the models
over the range $-2.5 \la \overH{Fe} \la -0.5$
\citep[and slightly overpredicted at lower
metallicities; see figure~1 of][]{francois-etal04}.
Interestingly, these are precisely the metallicities
where most of the CMB-regulated star formation occurs within our
simulations, and \isotope{24}{Mg} would indeed be produced at
greater rates with a top-heavy IMF. However, the O
abundances do not support this conjecture: they are, if anything
\textit{overpredicted} at these metallicities, while a top-heavy
IMF would have an even larger effect on \isotope{16}{O} than on
\isotope{24}{Mg}. Unless
some unaccounted-for process systematically drives down the \isotope{16}{O}
abundance by a significant amount, a more parsimonious explanation is
that the yields require adjustment, as suggested by \citet{francois-etal04}.

\section{Conclusions}%
\label{section:conclusions}

We have investigated the suggestion of \citet{larson05} and
\citet{stson} that star formation at high redshift and high
metallicity results in a top-heavy IMF due to the influence
of the CMB as an effective temperature floor for the collapsing
protostellar core. By assuming that CMB regulation is important
if the minimum temperature the core would reach in the absence
of the CMB, according to the models of \citet{omukai-etal05},
is less than the CMB temperature at a given redshift, we are
able to parametrize the evolution of the critical metallicity
\ZCMB\ as a function of redshift. Comparison with the 3D
hydrodynamical simulations of \citet{stson} suggests that our
model is generally accurate, but may underestimate \ZCMB\ by up to
an order of magnitude in particular cases.

By comparing \ZCMB\ to the metallicities of stars formed in
high resolution cosmological simulations of the formation of
eight $\sim L_*$ galaxies, we conclude that if CMB regulation does
operate, the majority of stars
at redshifts between $2.7$ and $6$ would have formed in the CMB-regulated
regime. We have investigated five possible observational signatures
of CMB regulation:
\begin{enumerate}
\item Current predictions of supernova rates, and possibly GRB rates,
are underestimates at $z \ga 3$.
\item Indirect measures of the star formation history based on the
stellar populations left behind by episodes of star formation, such
as the age distribution of local stellar populations and evolution
of the stellar mass density as a function of redshift, would systematically
underestimate the star formation rate at redshifts greater than $2.7$
compared to measurements based
on UV and far-infrared flux generated directly by star formation. Such
a discrepancy is seen, although it appears to extend to lower redshift.
\item Globular clusters that formed at high metallicity and high redshift
may have quickly evaporated, leaving no present-day evidence of their
existence, possibly explaining the relatively narrow metallicity
range of the metal-poor GC population. In this case, we would expect
there to be no extant GCs that are both metal-rich and formed at
very high redshift.
Because of the difficulties in determining absolute ages to old
GCs, and the steepness of the age-redshift relation at high redshift,
current data are not able to constrain this hypothesis.
\item Cosmic metallicity may rise more rapidly at early times than
predicted by current simulations that assume a normal IMF,
reaching solar metallicities at higher redshift. This could reconcile
the simulations with observations that show a mean metallicity of
approximately solar for stars formed out to redshift $3$.
\item Elements produced by high-mass supernova progenitors, particularly
the $\alpha$ elements O, Ne, and Mg, would be produced with
greater abundance between redshifts $6$ and $3$, or $\overH{Fe} \sim
-2$ to $-0.5$, than predicted in models that assume a normal IMF.
This could help explain the high Mg abundances seen
in stars of this metallicity range in the Galaxy, but would exacerbate
the overprediction of O over the same range in metallicities.
Since the O yields are more sensitive to the IMF, we
conclude that the abundance patterns are not produced by
a top-heavy IMF, although a bottom-light IMF would not alter
the abundances.
\end{enumerate}

In conclusion, the observations neither provide conclusive evidence
for the hypothesis that the CMB can regulate star formation,
nor rule it out.
The most convincing
evidence would come in the form of either the clear presence or
absence of an envelope in redshift-metallicity
space, above which stars are either absent or have greatly reduced
numbers compared to the number of stars believed to have formed there
from other measurements; unfortunately, as discussed in \S~\ref{section:gcs},
it is unclear if stellar ages will ever be known precisely enough
to perform this test.
Another potential avenue of exploration would be to use the methods
that currently argue for a mean change in the IMF with time,
such as in \citet{dave08} and \citet{wth08}, but subdividing the
populations by metallicity to determine if
an evolving IMF is only required for metal-rich stars, as would
be expected if the evolution in the IMF is due to CMB regulation,
or if it must be truly universal.

\acknowledgments
This paper makes use of simulations performed as part of the
SHARCNET Dedicated Resource project: ``MUGS: The
McMaster Unbiased Galaxy Simulations Project'' (DR316, DR401, and DR437).

\bibliography{../../masterref.bib}

\end{document}